\documentclass[sigplan,10pt]{acmart}
\renewcommand\footnotetextcopyrightpermission[1]{}
\AtBeginDocument{%
  }

\usepackage{xspace}
\newcommand{\sysname}{\textsc{xDSM}\xspace}

\usepackage{enumitem}
\usepackage{comment}
\usepackage{svg}
\usepackage{multirow}
\usepackage{pifont}
\usepackage{xcolor}
\definecolor{blue}{rgb}{0,0,0}
\definecolor{red}{rgb}{0,0,0}
\definecolor{green}{rgb}{0,0,0}

\begin{document}

\title{Scaling Unmodified Multithreaded Applications with Elastic CXL-based Distributed Shared Memory}

\author{Guowei Liu}
\affiliation{%
  \institution{Tianjin University}
  \city{Tianjin}
  \country{China}}
\email{guoweiu@tju.edu.cn}

\author{Kang Chen}
\affiliation{%
  \institution{Tsinghua University}
  \city{Beijing}
  \country{China}}
\email{chenkang@tsinghua.edu.cn}

\author{Laiping Zhao}
\affiliation{%
  \institution{Tianjin University}
  \city{Tianjin}
  \country{China}}
\email{laiping@tju.edu.cn}

\author{Yiming Li}
\affiliation{%
  \institution{Tianjin University}
  \city{Tianjin}
  \country{China}}
\email{l_ym@tju.edu.cn}

\author{Hanwen Liu}
\affiliation{%
  \institution{Tianjin University}
  \city{Tianjin}
  \country{China}}
\email{liuhanwen2024@tju.edu.cn}

\author{Chen Peng}
\affiliation{%
  \institution{Tianjin University}
  \city{Tianjin}
  \country{China}}
\email{pengchen513@tju.edu.cn}

\author{Yichi Chen}
\affiliation{%
  \institution{Tianjin University}
  \city{Tianjin}
  \country{China}}
\email{chenyichi@tju.edu.cn}

\author{Sheng Chen}
\affiliation{%
  \institution{Tianjin University}
  \city{Tianjin}
  \country{China}}
\email{chensheng@tju.edu.cn}

\author{Zhiyuan Su}
\affiliation{%
  \institution{Inspur Electronic Information Industry Co., Ltd.}
  \city{Jinan}
  \country{China}}
\email{suzhiyuan@inspur.com}

\author{Wenyu Qu}
\affiliation{%
  \institution{Tianjin University}
  \city{Tianjin}
  \country{China}}
\email{wenyu.qu@tju.edu.cn}

\renewcommand{\shortauthors}{Liu et al.}

\begin{abstract}

While CXL presents a promising hardware substrate for Distributed Shared Memory (DSM), seamlessly scaling multithreaded applications across multiple nodes remains a formidable challenge. Existing CXL-based DSMs fall short: they require manual code modifications to share non-heap data, employ rigid data placement policies that fail under diverse and dynamic workloads, and suffer from severe page-fault processing overheads in sub-microsecond ($\mu\mathrm{s}$) environments.

We present \sysname{}, a full-space, elastic DSM system built over CXL that transparently scales unmodified multithreaded applications. To eliminate the burden of manual code rewrites, \sysname{} employs an OS-runtime co-design that establishes a globally coordinated address space, seamlessly sharing all memory segments. To mask CXL access penalties, \sysname{} abandons static placement rules in favor of a dynamic, latency-driven policy that actively balances data between local DRAM and CXL memory. Finally, to resolve the fundamental tension between high base-page fault overheads and severe huge-page false sharing, \sysname{} introduces spatial locality-aware elasticity, dynamically coalescing and splitting pages on the fly to amortize processing costs. 

Evaluated across diverse workloads using 15 system configurations, \sysname{} outperforms CXL-only baselines by 1.5$\times$ to 2.2$\times$ and state-of-the-art hybrid DSMs by 1.1$\times$ to 2.2$\times$, while achieving near-linear scalability.
\end{abstract}

\settopmatter{printfolios=true,printacmref=false}
\maketitle
\pagestyle{plain}

\section{Introduction}

Data-intensive applications, such as graph analytics~\cite{Low2012GraphLab,Shun2013Ligra}, data mining~\cite{Zaharia2012Spark}, and machine learning~\cite{Abadi2016TensorFlow,Li2014ParameterServer}, place unprecedented demands on memory capacity and computational throughput. When applications outgrow the resources of a single machine, they must scale out across multiple nodes. However, manual porting of multi-threaded programs~\cite{posix1003} to distributed models, such as message passing or PGAS~\cite{dewael2015pgas,chapman2010openshmem}, is labor-intensive and error-prone. Distributed shared memory (DSM) systems~\cite{li1988ivy,keleher1994lazy,mueller1997dsmthreads} were designed to bridge this gap. They provide the abstraction of a single shared address space across multiple nodes with software-maintained memory consistency. Yet, their reliance on Ethernet or RDMA networks introduces substantial protocol overhead, typically slowing applications by an order of magnitude~\cite{ma2024drust,nelson2015grappa}. Compute Express Link (CXL) 3.0~\cite{cxl30whitepaper} provides a new opportunity to build efficient DSM systems by replacing slow software coherence with hardware-managed, multi-node cache coherence via {\color{blue}Back-Invalidation Snoop (BISnp)}.


Despite this powerful hardware capability, constructing an efficient full-space DSM across multiple nodes remains challenging. The \textbf{CXL-only} approach places all data on the shared CXL device. While avoiding software protocol overheads, it incurs 2.0$\times$ to 4.0$\times$ higher latency than local DRAM~\cite{sun2023demystifying,liu2025firework}, degrading performance by up to 2.2$\times$ (\S\ref{sec:eval-e2e}). Conversely, the \textbf{local-only} approach treats CXL merely as a communication channel and pulls all data into local memory~\cite{keleher1994lazy,keleher1992lrc}. This wastes CXL capacity and triggers frequent cross-node invalidations, causing up to an 8.0$\times$ performance degradation (\S\ref{sec:eval-e2e}). Firework~\cite{liu2025firework} introduces a \textbf{hybrid} design utilizing both tiers via a \emph{static} placement rule where shared pages map to CXL and private pages remain local. However, this static policy fails dynamic workloads. For instance, read-shared pages are forcibly placed on the CXL tier, inflicting severe latency penalties on every access.


This paper presents \sysname, a full-space, CXL-based \textbf{hybrid} DSM system. Through an OS-runtime co-design, \sysname bridges the low latency of local DRAM and the massive capacity of CXL memory. This architecture allows it to seamlessly execute unmodified multithreaded (\texttt{pthread}) programs across CXL-connected nodes. To realize this execution model, \sysname introduces three architectural innovations that directly overcome the fundamental limitations of existing systems.

First, existing DSM systems are fundamentally restricted by the \textbf{heap-only sharing limitation}. They rely on customized memory allocators that leave global and static variables unshared, forcing developers to manually rewrite code to achieve distributed execution. \sysname overcomes this by natively establishing a full-space shared execution environment. Through globally coordinated addressing and lightweight VMA(Virtual Memory Area)-anchored page transfers, the system shares all memory segments to eliminate manual code modifications entirely.

Second, current hybrid systems suffer from \textbf{rigid page placement limitations}. They typically enforce static rules, such as mapping all shared pages to CXL memory and private pages to local DRAM, which severely penalizes performance under dynamic or phase-changing workloads. \sysname abandons these static rules in favor of an adaptive data placement policy. By utilizing hardware-sampled, exponentially binned latency histograms, the system actively restores latency equilibrium between local DRAM and CXL memory to dynamically mask access penalties.

Finally, existing approaches are constrained by \textbf{fixed-granularity limitations}, which create a dilemma in sub-microsecond CXL environments. Operating at a 4\,KB base granularity triggers overwhelming software page fault processing overheads, whereas adopting 2\,MB huge pages exacerbates false sharing and wastes bandwidth. \sysname resolves this tension by introducing spatial locality-aware elasticity. The system dynamically coalesces contiguous pages in the background to amortize fault processing costs and performs on-demand splitting to isolate write-induced invalidations.

The novelty of \sysname lies in a comprehensive solution that leverages CXL shared memory to seamlessly scale multithreaded programs across nodes, while fundamentally eliminating the performance bottlenecks associated with rigid, fixed-granularity memory placement. We extensively evaluate the system using five diverse benchmarks. \sysname achieves up to a 2.2$\times$ speedup over CXL-only architecture, outperforms traditional local-only DSMs by up to 8.0$\times$, and surpasses static-hybrid DSMs by up to 2.2$\times$.
Furthermore, \sysname seamlessly adapts to dynamic workloads, achieving near-linear scalability across multiple nodes.

In summary, this paper makes the following contributions:
\begin{itemize} 
    \item We comprehensively analyze existing approaches, identifying three fundermental limitations: the lack full-space sharing, inflexible static data placement, and severe page-fault overheads.
    \item Our system, \sysname, is a full-space CXL-based hybrid DSM featuring three core innovations: Coordinated Addressing with VMA-anchored page transfer, Gap-Driven Asymmetric Migration for latency equilibrium, and Spatial Locality-Aware Elasticity.
    \item We thoroughly evaluate \sysname. The results demonstrate that \sysname significantly outperforms CXL-only, local-only, and static-hybrid DSMs.
\end{itemize}

\section{Background and Motivation}
\label{sec:background}

\begin{table}[t]
\centering
\caption{Comparison of DSM systems. ``Addr.'' specifies whether the system relies on heap-only sharing or achieves full-space shared execution environment. ``Place.'' denotes the use of static or dynamic page placement strategies. ``Gran.'' refers to fixed or elastic page management granularity.}
\label{tab:comparison}
\footnotesize
\setlength{\tabcolsep}{4pt}
\begin{tabular}{llccc}
\toprule
\textbf{Category} & \textbf{System} & \textbf{Addr.} & \textbf{Place.} & \textbf{Gran.} \\
\midrule
Local-only & IVY~\cite{li1988ivy}          & Heap & N/A    & Fixed \\
Local-only & TreadMarks~\cite{keleher1994lazy} & Heap & N/A & Fixed \\
Local-only & DSM-Threads~\cite{mueller1997dsmthreads} & Heap & N/A & Fixed \\
Local-only & DRust~\cite{ma2024drust}       & Heap & N/A    & Object \\
\midrule
CXL-only & Baseline                        & Full  & N/A    & N/A \\
\midrule
Hybrid & Firework~\cite{liu2025firework} & Heap & Static rules & Fixed \\
\midrule
Hybrid & \textbf{\sysname{}}         & \textbf{Full} & \textbf{Dynamic} & \textbf{Elastic} \\
\bottomrule
\end{tabular}
\end{table}

\subsection{CXL 3.0 and Multi-Node Shared Memory}
\label{sec:bg-cxl}

Compute Express Link (CXL)~\cite{cxl30whitepaper} extends node capacity by allowing direct \texttt{load}/\texttt{store} accesses to remote memory. In CXL 3.0, a single memory device can simultaneously connect 8--16 independent compute nodes~\cite{cxl30whitepaper,huang2025tigon} and enforce hardware cache coherence via \emph{back-invalidation} (BISnp)~\cite{cxl30whitepaper,jain2024cxl}. This hardware-managed architecture fundamentally eliminates the software consistency overheads inherent in traditional network-based DSMs (e.g., {\color{blue}2\,$\mu$s} for RDMA network~\cite{288627}). CXL reduces remote access latency to merely 200\,ns--400\,ns, approaching the $\sim$100\,ns of local DRAM~\cite{sun2023demystifying}.

However, CXL 3.0 lacks a native global virtual address space, as compute nodes run independent OSes with isolated page tables. Constructing a multi-node DSM over CXL requires establishing a \emph{full-space shared execution environment} to guarantee identical mappings across nodes, alongside \emph{distributed thread management} to execute unmodified applications.
Leveraging this hardware substrate, existing CXL-based DSM systems fall into three categories. The \textbf{CXL-only} approach allocates all data exclusively on the shared CXL device, eliminating software consistency overheads but penalizing every access with higher interconnect latency. Conversely, the \textbf{local-only} approach provisions all data within local memory, utilizing CXL strictly as a high-speed communication medium similar to classical network-based DSM architectures~\cite{li1988ivy,keleher1994lazy}. Finally, the \textbf{hybrid} approach integrates both memory tiers, attempting to optimize performance by placing data based on specific access patterns~\cite{liu2025firework}.

\subsection{Limitations and Challenges}
\label{sec:bg-limitation}

Table~\ref{tab:comparison} compares representative systems. Drawing from this comparison, we identify three fundamental limitations in existing approaches and outline the system-level challenges required to overcome them.

\subsubsection{Challenge 1: Establishing a Full-Space Shared Execution Environment}
\label{sec:bg-l1}

Existing DSM systems, including local-only~\cite{li1988ivy,keleher1994lazy,mueller1997dsmthreads,ma2024drust} and hybrid~\cite{liu2025firework} designs, rely on customized memory allocators to redirect heap allocations into shared memory. Consequently, global variables (\texttt{.data}), uninitialized globals (\texttt{.bss}), and function-scope statics remain unshared. Sharing these regions imposes a severe manual burden, as developers must explicitly rewrite static declarations into dynamic heap allocations and modify all corresponding access sites.


However, non-heap shared data is pervasive in real-world applications. Among the 13 PARSEC workloads~\cite{parsec}, 12 heavily rely on global or static variables for thread-shared states. For example, \texttt{fluidanimate} references 33 global variables (e.g., particle arrays and simulation parameters) across 401 lines in 9 files. Similarly, \texttt{vips} accesses 46 global variables at {\color{blue}1,843} reference sites scattered across 377 files. Manually converting such programs to a heap-only DSM is a prohibitively laborious and error-prone process.

To overcome this \textbf{heap-only sharing limitation}, a DSM must natively share all memory segments. The fundamental challenge is that existing OSes (e.g., Linux) lack native mechanisms to establish a globally shared address space across distributed nodes. Each node runs an independent kernel with its own VMAs and page tables, and runtime allocations independently assign virtual addresses that may easily conflict across nodes.

\subsubsection{Challenge 2: Adapting Data Placement to Dynamic Workloads}
\label{sec:bg-l2}

Page placement directly dictates application performance because local DRAM and CXL memory (\S\ref{sec:eval-setup}) differ significantly in latency (109 ns vs. 285 to 376 ns) and per-node bandwidth (36.5 GB/s vs. 29 to 37 GB/s).


Unfortunately, existing approaches adopt rigid page placement policies. The CXL-only approach forces all accesses through higher-latency links, whereas the local-only approach wastes CXL capacity and bandwidth. Firework~\cite{liu2025firework} attempts a hybrid design but maps shared pages to CXL and private pages locally. This static rule fails to accommodate dynamic workloads. For instance, forcing read-shared pages to CXL during a 16-thread BFS on a 6.8\,GB read-only graph degraded performance by 34\% ($2.41 \times 10^8$ vs.\ $3.25 \times 10^8$ traversed edges per second) compared to local replication.


To overcome this \textbf{rigid page placement limitation}, the system must replace static rules with an adaptive policy that distributes data across heterogeneous tiers on the fly. This is challenging because optimal placement depends on both the sharing pattern (e.g., private, shared-read, or shared-write) and the real-time load on each tier, both of which fluctuate significantly during runtime.

\subsubsection{Challenge 3: Mitigating Costly Page-Fault Processing Overheads}
\label{sec:bg-l3}

Existing DSM systems typically rely on page faults to transparently detect memory accesses and trigger data migration~\cite{li1988ivy,keleher1994lazy,mueller1997dsmthreads,liu2025firework}. Inherently, these systems adopt the fixed page granularity of the underlying hardware, typically 4\,KB (Table~\ref{tab:comparison}).


Traditional network-based DSMs tolerate page fault processing overheads as network delays dominate the critical path. 
Because CXL data accesses complete in under 400ns, the 24$\mu$s to 30$\mu$s software pipeline of a page fault (kernel trap, lock acquisition, and permission updates) instantly becomes a severe bottleneck~\cite{calciu2021kona}. This creates a strict fixed-granularity dilemma. Operating at a 4KB base page granularity triggers overwhelming trap overheads, whereas adopting 2MB huge pages reduces fault counts but exacerbates false sharing and wastes interconnect bandwidth~\cite{calciu2021kona}.

To overcome this \textbf{fixed-granularity limitation}, the system must redesign traditional page fault processing.
The core challenge is breaking this rigid granularity barrier to resolve the inherent tension between overwhelming fault processing costs at 4\,KB and severe false sharing at 2\,MB.

\subsubsection{Motivation: The Need for a New Comprehensive Solution} 
\label{sec:bg-summary}

To fully unleash the potential of CXL-based DSM and execute unmodified multithreaded programs efficiently, a new comprehensive solution is required. Specifically, the system must (1) natively establish a full-space shared execution environment without imposing manual code-rewrite burdens, (2) replace rigid rules with an adaptive policy to optimally distribute data across heterogeneous tiers, and (3) break the fixed-granularity barrier to amortize $\mu\mathrm{s}$-scale page fault processing overheads. 

\begin{figure*}[t]
  \centering
  \includegraphics[width=0.95\linewidth]{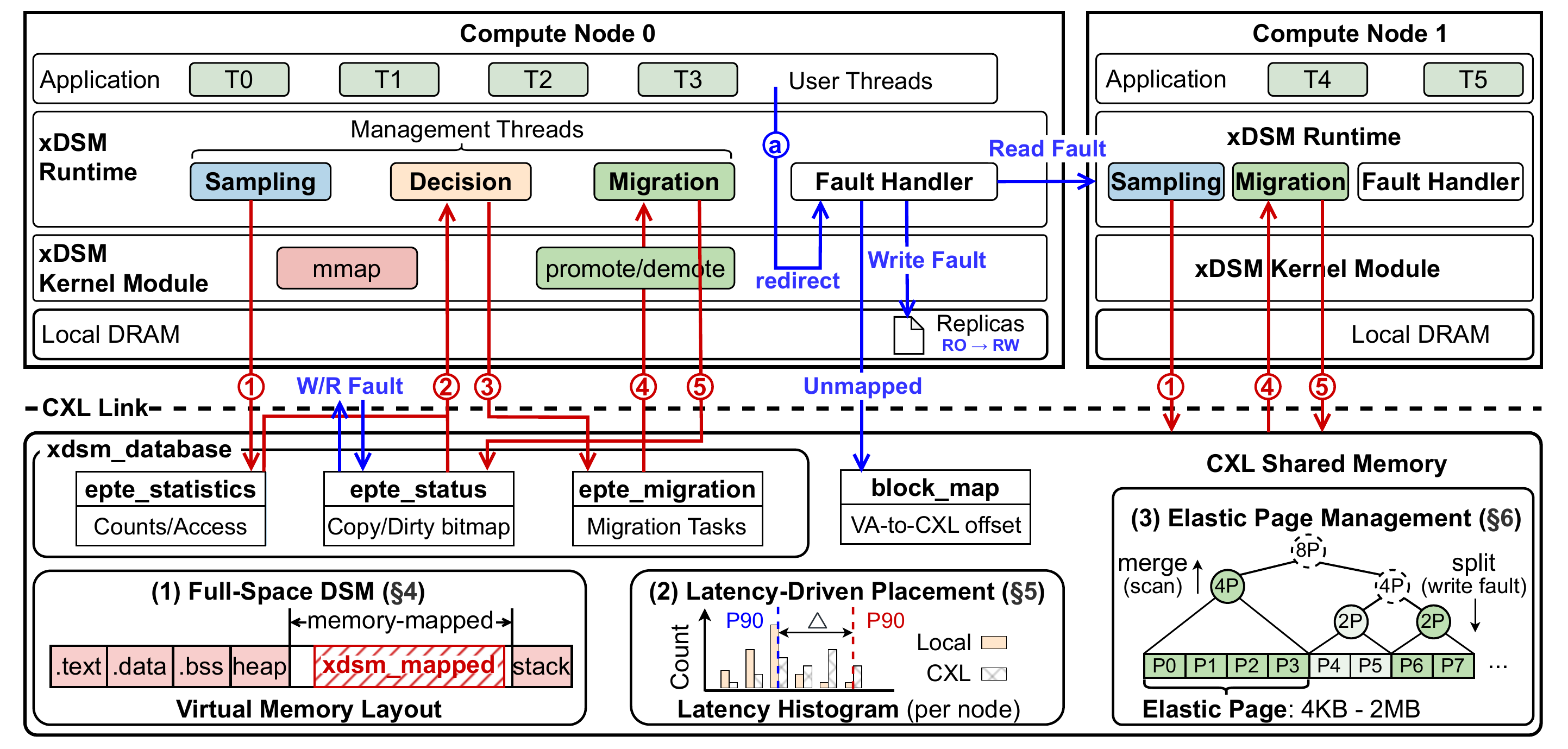}
  \caption{Overview of \sysname architecture. The system employs an OS-runtime co-design. In user space, three types of management threads (Sampling, Decision, and Migration) as well as page fault handlers coordinate memory migration and consistency. In the shared CXL memory, a global \texttt{xdsm\_database} maintains three core tables: \texttt{epte\_statistics} tracks memory access statistics (e.g., access counts), \texttt{epte\_status} manages consistency states (e.g., owner and dirty bitmaps), and \texttt{epte\_migration} queues tiering tasks. Additionally, a global \texttt{block\_map} records the deterministic mapping from virtual addresses to CXL offsets.}
  \label{fig:arch}
\end{figure*}

\section{\sysname Design}
\label{sec:overview}

\subsection{Architecture Overview}
\label{sec:arch-overview}

\sysname's overall architecture (Figure~\ref{fig:arch}) follows an OS-runtime co-design.
To efficiently manage sub-$\mu\mathrm{s}$ CXL memory, \sysname introduces several foundational abstractions. Spatially, the virtual address space is statically divided into fixed-size \textbf{Chunks} (e.g., 2\,MB) as the top-level allocation boundaries, while the OS natively manages 4\,KB \textbf{Base Pages} at the bottom. 
\textbf{Elastic Page} is \sysname's core management unit. 
It is an entity dynamically aggregated from contiguous base pages with {\color{red}same accessors}, varying in size from 4\,KB to 2\,MB. All metadata tracking and data migration are executed at this elastic granularity.

Temporally, \sysname operates in fixed-duration windows called \textbf{Epochs} (or \textbf{Ticks}, 1\,s by default). 
{\color{blue}During each tick, \textbf{Sampling Threads} profile memory accesses via hardware performance counters (one sample per 20,000 operations, where each operation is a macro-op dispatched after instruction decode). 
Each sampled L3-miss access is recorded in \texttt{epte\_statistics}. For load misses, the cache-miss latency is additionally recorded into a per-node histogram (\S\ref{sec:latency-metric}).
}
At the tick boundary, a singleton \textbf{Decision Thread} reads these statistics alongside the consistency states from \texttt{epte\_status} to generate batch page migration plans. These plans instruct the system to either \textbf{Promote} hot pages from the shared CXL device to local DRAM for latency reduction, or \textbf{Demote} local pages back to CXL to relieve memory pressure. The tasks are queued in \texttt{epte\_migration}, where per-node \textbf{Migration Threads} dequeue and execute them via lightweight kernel module interfaces.
Together, these coordinated abstractions and components resolve the three fundamental challenges, which we distill into the following architectural insights.

\subsubsection{Insight 1: Coordinated Global Addressing and Lightweight VMA-Anchored Page Transfers (\S\ref{sec:transparent})}
\label{sec:insight-fullspace}

To resolve heap-only sharing limitations (\textbf{C1}) and execute unmodified applications, \sysname establishes a globally coordinated virtual address layout. At initialization, it maps the entire application memory directly to the shared CXL device to guarantee a unified global view.
To prevent collisions during runtime, all local memory allocations are assigned from strictly disjoint virtual address ranges across nodes. Meanwhile, any memory mapped to the CXL device retains the exact same virtual address globally. Consequently, the system guarantees a single unified address space at all times.

\sysname further introduces a \emph{lightweight VMA-anchored page transfers} mechanism to accelerate memory tiering. {\color{blue}In Linux, a VMA, represented by \texttt{vm\_area\_struct}, describes a contiguous range of virtual addresses that have the same permission attributes and are backed by the same object (e.g., a file or anonymous mapping)~\cite{linux-mm-concepts}.} When moving data between tiers, \sysname updates only the underlying Page Table Entries (PTEs) to point to local memory. It deliberately leaves the VMA unmodified. The VMA acts as an anchor pointing to the CXL backing store. This decoupling makes promote and demote operations extremely lightweight on the fast path. It fundamentally avoids expensive cross-node coordination, severe \texttt{mmap\_lock} contention, and the overhead of repeatedly re-allocating shared CXL memory. Heavyweight VMA modifications are deferred to a slow path, invoked solely to unmap the CXL region and reclaim remote capacity when a page achieves long-term local residency.

\subsubsection{Insight 2: Restoring Latency Equilibrium via Gap-Driven Asymmetric Migration (\S\ref{sec:placement})}
\label{sec:insight-latency}

To overcome the limitations of static placement rules and dynamically adapt to shifting workloads (\textbf{C2}), \sysname abandons rigid policies in favor of an adaptive placement policy. Existing tiered memory systems often rely on average access latencies~\cite{vuppalapati2024colloid,liu2025alto}, a single-value metric highly vulnerable to long-tail outliers. \sysname addresses this limitation by introducing \emph{Exp-binned Latency Histograms}. By recording hardware-sampled latencies into exponentially growing intervals, \sysname naturally absorbs long-tail anomalies and establishes a stable metric. The placement policy's core objective is to restore and maintain \emph{latency equilibrium}, which is defined as \emph{P90-Bin Alignment}---the state where the 90th-percentile latency bins of local DRAM and CXL memory intersect.

To swiftly achieve this equilibrium, \sysname pipelines the execution through two mechanisms. First, it employs \emph{Gap-Proportional Volume Scaling}, dynamically computing the number of pages to migrate proportional to the measured P90-bin gap ($\Delta$). Second, it enforces \emph{Asymmetric Migration Pacing} during execution. \sysname aggressively promotes active pages to mask CXL latency penalties, while conservatively throttling demotion volumes to prevent thrashing.

\subsubsection{Insight 3: Spatial Locality-Aware Elasticity (\S\ref{sec:elastic})}
\label{sec:insight-elasticity}

To resolve the fundamental trade-off between the high trap overhead of 4\,KB base pages and the false sharing exacerbated by 2\,MB huge pages (\textbf{C3}), \sysname overcomes this fixed-granularity limitation via \emph{spatial locality-aware elasticity}. 
Instead of enforcing a static page size, 
\sysname dynamically merges contiguous pages with identical consistency states and {\color{blue}the same accessors} into variable-sized elastic pages (up to 2,MB)
This background coalescing significantly amortizes fault resolution and migration overheads. 
Conversely, since a write fault on a shared large elastic page would invalidate the entire region and penalize unrelated sub-regions, \sysname performs on-demand splitting upon write faults, iteratively reducing
the page size down to 4\,KB. This approach isolates write-induced invalidations, eliminating false sharing penalties while retaining the efficiency of bulk migration.

\section{Full-space Shared Environment}
\label{sec:transparent}
\label{sec:design}

\begin{table}[t]
  \centering
  \caption{Tables in \texttt{xdsm\_database}.}
  \label{tab:epte-fields}
  \footnotesize
  \setlength{\tabcolsep}{4pt}
  \renewcommand{\arraystretch}{0.92}
  \begin{tabular}{@{}p{0.32\columnwidth}p{0.68\columnwidth}@{}}
    \toprule
    \textbf{Table / Field}           & \textbf{Description}                   \\
    \midrule
    \multicolumn{2}{@{}l}{\textbf{\texttt{epte\_status}}}                    \\
    \midrule
    \texttt{private}                 & Exclusive ownership flag.              \\
    \texttt{owner}                   & Owner node ID.             \\
    \texttt{dirty}                   & Bitmask of nodes with modified copies. \\
    \texttt{copyset}                 & Bitmask of nodes holding local copies. \\
    \texttt{lock}                    & Per-entry mutex.                       \\
    \midrule
    \multicolumn{2}{@{}l}{\textbf{\texttt{epte\_statistics}}}                \\
    \midrule
    \texttt{ld\_cnt/st\_cnt}         & Per-tick load/store counts.            \\
    \texttt{\color{red}{accessor}}                  & Per-tick accessor node bitmask.        \\
    \texttt{hist\_\{ld,st\}\_\color{red}{accessor}} & Cumulative reader/writer node bitmask (periodically reset). \\
    \midrule
    \multicolumn{2}{@{}l}{\textbf{\texttt{epte\_migration}}}                 \\
    \midrule
    \texttt{vaddr/nr\_pages}         & Target address range.                 \\
    \texttt{type}                    & Transfer type (promote-move/ promote-copy /demote).      \\
    \texttt{target}                  & Nodes to execute the transfer.         \\
    \bottomrule
  \end{tabular}
\end{table}

\subsection{Coordinated Global Addressing}
\label{subsec:space_management}

\noindent \textbf{Preserving a Unified Address Space.}
The fundamental goal of \sysname is to project a globally identical virtual address (VA) space across all distributed threads. At initialization, \sysname maps the entire application memory layout (including \texttt{.text}, \texttt{.data}, \texttt{.bss}, \texttt{heap}, {\texttt{memory-mapped},} and \texttt{stack}) directly to the shared CXL device. This guarantees a uniform VA projection across all nodes at startup (Figure~\ref{fig:arch}). During runtime, driven by the adaptive placement policy (\S\ref{sec:placement}), actively accessed pages are dynamically promoted to local memory, creating a hybrid backing store. To preserve global VA uniformity under this dynamic model, \sysname enforces a strictly coordinated allocation scheme. 
First, all local non-shared memory allocations are assigned from strictly disjoint VA ranges across nodes to prevent collisions. 
Second, any memory mapped to the CXL device is mapped to the exact same VAs globally. 
Consequently, a global unified address space is guaranteed, allowing native execution of unmodified applications.

\noindent \textbf{CXL Memory Management.}
{Within the \texttt{memory-mapped} segment}, \sysname reserves an \texttt{xdsm\_mapped} region backed by CXL physical memory. \sysname implements a two-level memory allocator here. A global block allocator manages 64\,MB blocks via a global bitmap and \texttt{block\_map}, while a per-node allocator handles fine-grained allocations. Crucially, this region also hosts the globally shared \texttt{xdsm\_database} (Table~\ref{tab:epte-fields}), which underpins the latency-driven placement policy (\S\ref{sec:placement}) and elastic page management (\S\ref{sec:elastic}).

\subsection{Lightweight VMA-Anchored Page Transfers}
\label{sec:coherence}

\noindent \textbf{Decoupled VMA and Page Tables.}
Each node maintains its own local Virtual Memory Area (VMA) structures and page tables. To ensure high performance when transferring data between tiers, \sysname introduces \emph{VMA-anchored fast paths}. When promoting or demoting a page, \sysname updates only the underlying Page Table Entries (PTEs) to point to local memory, leaving the VMA unmodified. The VMA acts as an anchor pointing to the CXL backing store. This decoupling makes operations lightweight on the fast path because it fundamentally avoids expensive cross-node coordination. Specifically, a node does not need to wait for other nodes to unmap their regions before it can update its own mapping to local memory. Furthermore, this approach eliminates the overhead of repeatedly freeing and re-allocating shared CXL memory during dynamic tiering. Heavyweight VMA modifications are deferred to a slow path, invoked solely to detach the CXL connection and reclaim remote capacity when a page achieves long-term local residency.

\noindent \textbf{Fast-Path Page Transfers.}
At the end of each tick, the decision thread reads access statistics from \texttt{epte\_statistics} and consistency states from \texttt{epte\_status} (\ding{173} in Figure~\ref{fig:arch}), then categorizes each elastic page into one of three fast-path transfer actions, enqueued in \texttt{epte\_migration} (\ding{174}):
\begin{itemize}[leftmargin=*,nosep]
  \item \textbf{Promote-Move:} For pages with only one accessor {(single bit in \texttt{\color{red}accessor} and \texttt{hist\_\{ld,st\}\_{\color{red}accessor}})}, \sysname moves the pages from CXL to that node's local memory for exclusive access {(\texttt{private} and \texttt{owner} will be set)}.
  \item \textbf{Promote-Copy:} For pages read by multiple nodes without writes {(multiple bits in \texttt{\color{red}accessor} with zero \texttt{hist\_st\_{\color{red}accessor}})}, \sysname replicates the pages to local memory of all readers as read-only copies {(\texttt{copyset} will be set)}.
  \item \textbf{Demote:} Local replicas on nodes under memory pressure are evicted back to CXL memory {(\texttt{private} and \texttt{copyset} will be cleared)}.
\end{itemize}
Per-node migration threads dequeue these tasks (\ding{175}) and execute the data transfers via the kernel module (\ding{176}).
Crucially, pages written by multiple nodes {(multiple bits in \texttt{hist\_st\_{\color{red}accessor}})} remain pinned in CXL memory to avoid costly consistency traffic.

\noindent \textbf{Transparent Access Resolution.}
Following page transfers, \sysname relies on page faults to transparently resolve memory accesses and enforce both Sequential Consistency (SC)~\cite{li1988ivy} and Release Consistency (RC)~\cite{keleher1994lazy}.
The fault handler uses the \texttt{block\_map} and \texttt{epte\_status} to resolve accesses at elastic granularity on the critical path (\textcircled{a} in Figure~\ref{fig:arch}):
\begin{itemize}[leftmargin=*,nosep]
  \item \textbf{Unmapped Fault:} Triggered when accessing an unmapped page. The handler resolves the mapping via \texttt{block\_map} and establishes the VMA mapping.
  \item \textbf{Read Fault:} Triggered when reading an invalidated page, e.g., the page is exclusively owned by another node. The handler forces the owner to flush dirty data to CXL, then grants read-only (\texttt{RO}) {\color{red}accessor} to intercept future writes.
  \item \textbf{Write Fault:} Triggered when writing to a read-only page. Under SC, the handler invalidates all remote copies {(identified by \texttt{copyset})} before granting read-write (\texttt{RW}) {\color{red}accessor}. Under RC, it upgrades to \texttt{RW} locally and defers invalidations until the next synchronization point.
\end{itemize}

To prevent race conditions when background page transfers interleave with concurrent page fault handling, \sysname strictly coordinates these events using the fine-grained \texttt{lock} embedded within each \texttt{epte\_status} entry.

\subsection{Application Thread Management}
\label{sec:dist-thread}

\sysname intercepts POSIX threading APIs to transparently support unmodified distributed execution.

\noindent \textbf{Thread Lifecycle.}
The runtime intercepts \texttt{pthread\_create} to dispatch execution to remote nodes via a round-robin policy. Because the VA space is globally uniform, function pointers and arguments remain natively valid cluster-wide. Each thread is assigned a globally unique identifier (GUID), which the runtime uses to transparently track and manage remote threads during operations such as \texttt{pthread\_join}.

\noindent \textbf{Synchronization.}
Standard synchronization primitives (e.g., mutexes, spinlocks, barriers) are natively supported by allocating their underlying objects directly within the shared CXL memory. The user-space runtime intercepts these primitives to seamlessly enforce the configured consistency model without requiring source-code modifications.

\begin{figure}[t]
  \centering
  \includegraphics[width=\columnwidth]{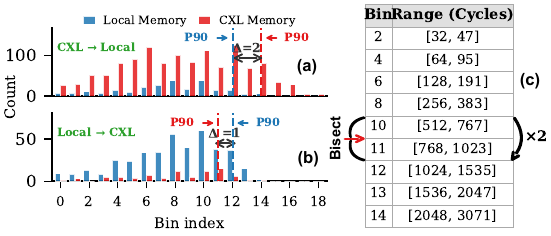}
  \caption{
  Per-node latency histograms for local (blue) memory and CXL (red) from Jacobi (\S\ref{sec:eval-bench}). Dashed lines mark P90 bins. (a) CXL P90 exceeds local P90, triggering promotion. (b) After promotion saturates local DRAM, local P90 exceeds CXL P90, triggering demotion. 
  (c) Bin-to-latency mapping. {\color{blue}Intervals $[2^k, 2^{k+1})$ grow by powers of two and each is bisected into two bins: e.g., bin~10 and bin~11 split the interval $[512, 1024)$ into two equal-width bins $[512, 767]$ and $[768, 1023]$, while bin~12 begins the next interval $[1024, 2048)$.}
  }
  \label{fig:lat-hist}
\end{figure}

\section{Latency-Driven Page Placement}
\label{sec:placement}


Effective page placement is critical to tiered-memory DSM performance. Existing hybrid systems often rely on rigid static rules~\cite{liu2025firework} or a single value such as average access latency~\cite{vuppalapati2024colloid,liu2025alto} to guide data movement. Rigid static rules fail to accommodate diverse access patterns. Average latency metrics are vulnerable to outliers in sub-$\mu\mathrm{s}$ CXL environments, causing placement instability. To overcome these limitations, \sysname employs an adaptive, latency-driven placement policy. This policy maintains a latency equilibrium between local memory and CXL memory.

\subsection{Exp-Binned Latency Histograms}
\label{sec:latency-metric}

To capture dynamic tier latencies and filter out long-tail anomalies, \sysname introduces \emph{Exp-binned Latency Histograms}. Sampling threads (\ding{172} in Figure~\ref{fig:arch}) continuously record L3 load miss latencies into two per-node histograms (one for local memory and one for CXL). 
{Because latencies within a workload exhibit high variance, the histograms categorize sampled latencies into exponentially growing intervals that compress outliers into fewer bins. Each interval is bisected into two equal-width bins to provide sufficient resolution for distinguishing local memory from CXL latency. For a latency sample $\ell$ (in cycles), its bin index is calculated by Equation~\ref{eq:bin}:}
\begin{equation}
  \label{eq:bin}
  \mathrm{bin}(\ell)=
  \begin{cases}
    0, & 1 \le \ell \le 15,\\
    1, & 16 \le \ell \le 31,\\
    2\!\left(\lfloor \log_2 \ell \rfloor - 4\right)
    + \left\lfloor
      \dfrac{\ell - 2^{\lfloor \log_2 \ell \rfloor}}
            {2^{\lfloor \log_2 \ell \rfloor - 1}}
      \right\rfloor,
    & \ell \ge 32.
  \end{cases}
\end{equation}


{The placement policy uses \emph{P90-Bin Alignment} as its criterion for latency equilibrium. At each tick, the decision thread computes the 90th-percentile (P90) bin index for local memory ($\textit{local\_bin}_i$) and CXL ($\textit{cxl\_bin}_i$) on each node $i$ (Figure~\ref{fig:lat-hist}). The P90 bin is the lowest index encompassing at least 90\% of samples. If $\textit{local\_bin}_i < \textit{cxl\_bin}_i$, the policy promotes active pages to local memory. If $\textit{local\_bin}_i > \textit{cxl\_bin}_i$, it demotes pages to CXL. If the two align, latency equilibrium is achieved and no migration is triggered.}

\subsection{Gap-Proportional Volume Scaling}
\label{sec:migration-volume}


Once the migration direction is determined, \sysname dynamically scales the migration volume in proportion to the measured P90-bin gap ($\Delta_i = |\textit{cxl\_bin}_i - \textit{local\_bin}_i|$). This gap reflects the severity of the performance imbalance. At each tick, the decision thread computes the migration volume for node $i$ by Equation~\ref{eq:volume}.
\begin{equation}
  \label{eq:volume}
  \textit{volume}_i = \frac{\textit{cnt}_i}{2^{\alpha - \min(\alpha, \Delta_i)}}
\end{equation}

{\color{blue}Here, $\textit{cnt}_i$ is the number of L3-miss samples observed on the source tier of node $i$ during the current tick,}
and $\alpha$ is a scaling parameter that controls migration aggressiveness (\S\ref{sec:asymmetric-pacing}). The gap $\Delta_i$ shrinks the denominator, increasing migration volume until it saturates at $\textit{cnt}_i$ when $\Delta_i \ge \alpha$. The decision thread then sorts elastic pages by per-tick access count (\texttt{ld\_cnt}+\texttt{st\_cnt}) in descending order, selecting pages for migration and decrementing $\textit{volume}_i$ by their access counts until $\textit{volume}_i$ reaches zero.

\subsection{Asymmetric Migration Pacing}
\label{sec:asymmetric-pacing}


{To address asymmetric CXL latencies and prevent placement oscillation, \sysname enforces \emph{Asymmetric Migration Pacing} by configuring $\alpha$ differently for promotion and demotion.}




For page promotion, \sysname acts aggressively by setting $\alpha = 2$ and using CXL sample count for $\textit{cnt}_i$ in Equation~\ref{eq:volume}. It rapidly moves hot pages to local memory to reduce CXL latency penalties. Conversely, for page demotion, \sysname acts conservatively by setting $\alpha = 4$ and using local memory sample count for $\textit{cnt}_i$. This higher $\alpha$ intentionally throttles the eviction rate. It bounds the demotion volume even when the latency gap is large, effectively preventing thrashing and stabilizing the local working set.
\section{Elastic Page Management}
\label{sec:elastic}

\subsection{Spatial Locality-Aware Elasticity}
\label{sec:grouping}

To resolve the fundamental trade-off between high base-page fault processing overheads and severe huge-page false sharing, \sysname introduces \emph{spatial locality-aware elasticity}. This mechanism dynamically coalesces contiguous base pages into variable-sized elastic pages ranging from 4\,KB to 2\,MB. By treating these coalesced regions as a single unified management entity, \sysname significantly amortizes the costs of fault resolution, permission updates, and page transfers.



The runtime governs this elasticity by exploiting spatial locality. Two adjacent elastic pages are eligible for coalescing only if they share the same consistency state in \texttt{epte\_status} (i.e., identical copyset, dirty bitmask, and owner, with no pending locks or in-flight transfers) and have {\color{blue}the same accessors} in \texttt{epte\_statistics}: {\color{blue}neither is written by multiple nodes, single-writer pages share same writer, and read-only pages share the same set of readers. Pages without any access can be merged with adjacent elastic pages.}

\subsection{Hierarchical Coalescing and Splitting}
\label{sec:adaptive-granularity}

To track these variable-sized pages with low metadata overhead, \sysname maintains a hierarchical bitmap within each 2\,MB chunk (Figure~\ref{fig:arch}~(3)). Level $e$ manages aligned groups of $2^e$ base pages, and a set bit indicates that the corresponding $2^e$ pages form a single elastic page. For instance, if an entire 512-page chunk exhibits uniform access behavior, it coalesces to level 9, represented by a single bit and enabling 2\,MB batch operations.

\noindent\textbf{Background Coalescing.}
At the end of each tick, immediately prior to making placement decisions (\S\ref{sec:placement}), the decision thread sweeps the bitmap to identify eligible sibling groups. If two adjacent level-$e$ siblings meet the coalescing criteria above, they are coalesced into a level-$(e{+}1)$ group. The bitmap is atomically updated by clearing the two child bits and setting the parent bit. This proactively enlarges the management granularity for read-heavy or stable memory regions, optimizing them for bulk transfers.

\noindent\textbf{On-Demand Splitting.}
On-demand splitting is triggered upon a write access to a read-only replica. When a write fault occurs on a shared elastic page, the fault handler triggers a split. It iteratively splits the covering level-$e$ group down to the 4\,KB base granularity. At each division step, the parent consistency state is propagated to both children. However, the bitmap traversal only continues splitting the specific half encompassing the faulting address. Consequently, write-induced invalidations are isolated to the targeted 4\,KB page. This leaves the remainder of the original elastic page intact and valid for remote nodes, eliminating false sharing penalties while retaining the efficiency of bulk transfers.

\section{Implementation}
\label{sec:impl}

We implemented \sysname on Linux 6.3 (x86-64) with $\sim$23,800 lines of C code: $\sim$21,400 in the user-space runtime (\texttt{libxdsm.\allowbreak{}so}) and $\sim$2,400 in the kernel module (\texttt{/dev/xdsm}). An additional $\sim$8,300 LOC implement other systems for comparison. 
{\color{blue}All source code will be open-sourced upon publication.}

\noindent\textbf{User-Space Runtime.} 
Injected via \texttt{LD\_PRELOAD}, the runtime intercepts \texttt{\_\_libc\_start\_main}, POSIX threading APIs, and memory allocators. At startup, it parses \texttt{/proc/self/\allowbreak{}maps}, reads existing pages via \texttt{process\_vm\_readv} and writes them to shared CXL memory, and logs VA-to-CXL mappings so remote nodes can reconstruct the same address space. 
Cross-node coordination leverages per-thread CXL ring buffers. 
{\color{blue}The runtime also implements both RC and SC.}

\noindent\textbf{Hardware Profiling.} 
Each node's sampling thread profiles memory accesses via AMD IBS~\cite{drongowski2007ibs}, opening one \texttt{perf\_event} descriptor per user thread (\texttt{l3missonly} mode) and polling ring buffers to harvest latency samples with low overhead.

\noindent\textbf{Kernel-Level Page Management.} 
The module exposes an \texttt{ioctl} interface for fast-path memory operations that modify only Page Table Entries (PTEs), leaving VMAs intact. 
The user-space runtime resolves all data races before these calls.
\begin{itemize}[leftmargin=*,nosep]
  \item \textbf{Promotion (\texttt{xdsm\_promote\_sharedpages}):} 
  It has three stages: snapshotting file-backed PTEs, allocating local anonymous pages to copy from CXL (\texttt{copy\_mc\_\allowbreak{}user\_highpage}), and {\color{blue}validating PTEs are unchanged since the snapshot before batch-updating mappings.}
  Both the snapshot and update stages process PTEs in batches under Page Table Locks (PTLs), with the final stage issuing a single \texttt{flush\_tlb\_range} for the entire batch.
  \item \textbf{Demotion (\texttt{xdsm\_demote\_sharedpages}):}
  It discards clean local pages. Otherwise, it writes back to the CXL pages via the VMA's file mapping (\texttt{filemap\_grab\_folio}) and frees memory using \texttt{MADV\_DONTNEED}.
\end{itemize}

\section{Evaluation}
\label{sec:eval}



\begin{figure*}[t]
  \centering
  \Description{End-to-end application performance with 32 threads (4 nodes).}
  \includegraphics[width=\textwidth]{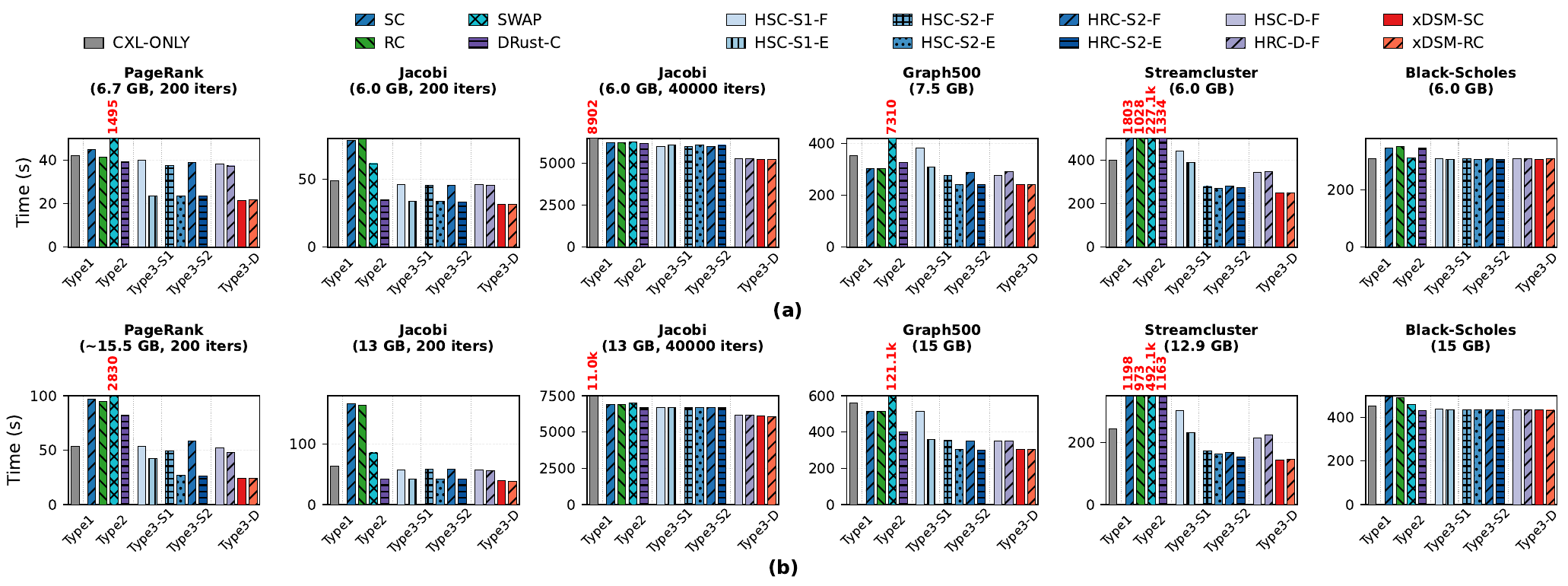}
  \caption{End-to-end execution time.(a)  16 threads (2 nodes), small datasets. (b) 32 threads (4 nodes), medium datasets.}
  \label{fig:app}
\end{figure*}


{We evaluate \sysname along four dimensions: (1)~end-to-end application performance across 15 configurations, (2)~per-iteration breakdown to evaluate the impact of placement strategy and page granularity separately, (3)~dynamic workload adaptation, and (4)~scalability from 1 to 4 nodes.}

\subsection{Experimental Setup}
\label{sec:eval-setup}

\noindent\textbf{Hardware.}
All experiments run on a dual-socket AMD EPYC server with Linux 6.3.
We enable NPS2 (NUMA-Per-Socket = 2) in BIOS, which partitions each socket into two NUMA domains, yielding four NUMA nodes. Each node has 8 physical cores and 4\,GB of local DDR5 DRAM (16\,GB aggregated). A 64\,GB CXL 1.1 Type-3 memory expander is attached and is accessible from all four NUMA nodes as a shared DAX device. Although CXL 3.0 multi-headed devices are not commercially available, CXL 1.1 supports hardware cache coherence within a single machine, satisfying the requirements of our systems. 
We use four NUMA nodes to emulate four independent compute nodes in a CXL-connected cluster, {\color{blue}this setting aligned with existing studies~\cite{alverti2025cxlfork,liu2025firework,huang2025tigon}.}

\noindent\textbf{Comparison systems.}
We compare 15 system configurations organized into three categories:

\noindent
\textbf{Type\,1: CXL-ONLY (baseline).}
All application data resides on the shared CXL device.

\noindent
\textbf{Type\,2: Local-only DSM.}
Application data is allocated on local DRAM and consistency is maintained through software protocols, using CXL only as the communication channel (replacing Ethernet/RDMA).
This category includes four systems adapted to our CXL platform:
\textbf{SC}~(Sequential Consistency, IVY~\cite{li1988ivy}),
\textbf{RC}~(Release Consistency, TreadMarks~\cite{keleher1994lazy}),
\textbf{SWAP}~(a page-swapping DSM that migrates full pages on access, similar to Fastswap~\cite{amaro2020far}), and
\textbf{DRust-C}~(a C reimplementation of DRust's~\cite{ma2024drust} ownership-based protocol).

\noindent
\textbf{Type\,3: Hybrid (local + CXL).}
These systems simultaneously use local DRAM and CXL memory with different \emph{placement strategies} and \emph{page granularities}.
We use the naming convention \textbf{H\textit{[consistency]}-\textit{[strategy]}-\textit{[granularity]}}, where \textbf{H} stands for Hybrid,  consistency is \textbf{RC} or \textbf{SC}, strategy is \textbf{S1/S2/D} (explained below), and granularity is \textbf{F}~(fixed 4\,KB page) or \textbf{E}~(elastic 4\,KB--2\,MB).

\begin{itemize}[nosep,leftmargin=*]
\item \textbf{Strategy\,1 (S1):} A static rule that classifies pages as shared or private. Shared pages remain on CXL, while private pages are promoted to local DRAM. This yields \textbf{HSC-S1-F} and \textbf{HSC-S1-E}.

\item \textbf{Strategy\,2 (S2):} A finer static rule that further distinguishes read vs.\ write. \emph{Private read-write} pages and \emph{shared read-only} pages are promoted to local DRAM, while \emph{shared read-write} pages remain on CXL. This yields \textbf{HSC-S2-F}, \textbf{HSC-S2-E}, \textbf{HRC-S2-F}, and \textbf{HRC-S2-E}.

\item \textbf{Strategy\,3 (D):} The latency-driven dynamic placement (\S\ref{sec:placement}).
  This yields \textbf{HSC-D-F}, \textbf{HRC-D-F}, and the two \sysname configurations: \textbf{\sysname{}-SC} (= \textbf{HSC-D-E}) and \textbf{\sysname{}-RC} (= \textbf{HRC-D-E}).
\end{itemize}

Type\,3 systems are compared against Type\,1 and Type\,2 to demonstrate the benefit of hybrid memory utilization.
The comparison of \textbf{S1}, \textbf{S2}, and \textbf{D} shows the effect of different placement strategies, while the comparison of \textbf{F} and \textbf{E} shows the effect of elastic page management.



\subsection{Benchmarks and Datasets}
\label{sec:eval-bench}

We select five applications with diverse access patterns:

\begin{itemize}[nosep,leftmargin=*]
\item \textbf{PageRank}~\cite{gap}: Iterative graph ranking over an adjacency matrix.
  Each thread computes ranks for a partition of vertices, reading the shared matrix and writing private rank vectors.
  Access pattern: dominated by \emph{private read-write} to rank arrays with read-only sharing of the matrix.

\item \textbf{Jacobi}~\cite{pouchet2016polybench}: 
   Iterative 2D stencil relaxation.
  Threads operate on horizontal strips and synchronize via barriers each iteration.
  Access pattern: predominantly \emph{private read-write} with boundary sharing between adjacent strips.

\item \textbf{Graph500}~\cite{graph500}: Breadth-first search on a large random graph stored in compressed sparse row (CSR) format.
  Multiple BFS traversals run in parallel from independent roots, with each thread reading the shared graph structure and maintaining a private visited set and queue.
  Access pattern: dominated by \emph{shared read-only} traversal of the CSR arrays with irregular, pointer-chasing access.
  We use the BFS kernel of Graph500 with our own pthread-based implementation.

\item \textbf{Streamcluster}~\cite{parsec}: An online clustering algorithm from the PARSEC benchmark suite.
  Threads cooperatively compute facility-location costs over a shared point set, with a mix of shared reads (point coordinates) and private writes (membership updates, cost accumulators).
  Access pattern: \emph{mixed} private read-write and shared read.

\item \textbf{Blackscholes}~\cite{parsec}: European option pricing from the PARSEC benchmark suite.
  Threads independently compute prices for disjoint partitions of the option array with no inter-thread sharing.
  Access pattern: \emph{compute-intensive} with small memory pressure.
\end{itemize}

\noindent\textbf{Dataset Scales.}
We consider three dataset scales:  
\emph{Small} datasets (4–8\,GB) for 2-node configurations.  
\emph{Medium} datasets (8–16\,GB) for 4-node configurations.  
\emph{Large} datasets (>16\,GB), exceeding four nodes' aggregated local DRAM, thus requiring CXL memory.
The data capacity of traditional DSM systems is strictly bounded by the aggregate local memory across all nodes. Consequently, Type 2 systems cannot execute workloads with large datasets.

\subsection{Application Performance}
\label{sec:eval-e2e}

{\color{blue}Figure~\ref{fig:app} compares} the end-to-end execution time of all 15 system configurations on 2 nodes (16 threads, small datasets) and 4 nodes (32 threads, medium datasets), respectively.
We analyze the results along four dimensions.

\noindent\textbf{Hybrid (Type\,3) vs.\ CXL-only (Type\,1).}
Across all memory-intensive benchmarks, the best hybrid configuration (\sysname{}-RC or \sysname{}-SC) outperforms CXL-ONLY by $1.5\times\text{--}2.2\times$.
At 2 nodes, \sysname{} reduces execution time by 49\% on PageRank, 36\% on Jacobi, 38\% on Streamcluster, and 32\% on Graph500.
Results of 4 nodes show even more advantage: 55\% on PageRank, 38\% on Jacobi, 41\% on Streamcluster, and 46\% on Graph500.
Hybrid systems place hot pages on local DRAM with lower latency, while CXL-ONLY forces all accesses through higher-latency CXL links.

\noindent\textbf{Hybrid (Type\,3) vs.\ local-only DSM (Type\,2).}
\sysname{}-RC outperforms the best Type\,2 system by $1.8\times$ on PageRank and by $8\times$ on Streamcluster at 4 nodes.
SWAP suffers from page thrashing on shared pages, as each shared page bounces between the two accessing nodes, making it orders of magnitude slower on Streamcluster.
The hybrid approach avoids this overhead: shared-read pages are replicated locally, and only write-contended pages incur consistency costs.

\noindent\textbf{Strategy comparison (S1, S2, D).}
Within Type\,3, more sophisticated placement strategies yield consistent gains.
Comparing at fixed-page granularity (-F) on 2 nodes: S1 places shared pages on CXL indiscriminately, so HSC-S1-F performs close to CXL-ONLY.
S2 improves by promoting shared-read pages to local DRAM, reducing execution time by 6\%--29\% over S1.
The dynamic strategy D further improves over S2 for long-running workloads: on Jacobi with 40\,000 iterations, D is 12.1\% faster than S2, because the latency-driven policy adapts to phase changes that static rules miss.

\noindent\textbf{Granularity comparison (F vs.\ E) at fixed strategy.}
At every strategy level, elastic groups (-E) outperform fixed pages (-F).
The benefit is most pronounced for PageRank: at S2 on 2 nodes, -E is 37\% faster than -F. At D on 4 nodes, \sysname{}-RC is 50\% faster than HRC-D-F.
Jacobi shows a similar pattern: 31\% faster at both 2 and 4 nodes.
The advantage comes from coalescing: PageRank and Jacobi access large contiguous arrays, so elastic groups batch up to 512 pages per fault, cutting fault count by two orders of magnitude.
Blackscholes is compute-bound and insensitive to memory placement: Type\,1 and Type\,3 configurations perform within 5\% of each other, while Type\,2 systems incur ${\sim}$12\% overhead from software consistency.

\subsection{Performance Breakdown}
\label{sec:eval-breakdown}

\begin{figure*}[t]
  \centering
  \Description{Per-iteration throughput breakdown for the iterative program (private read-write intensive, 16 threads).}
  \includegraphics[width=\textwidth]{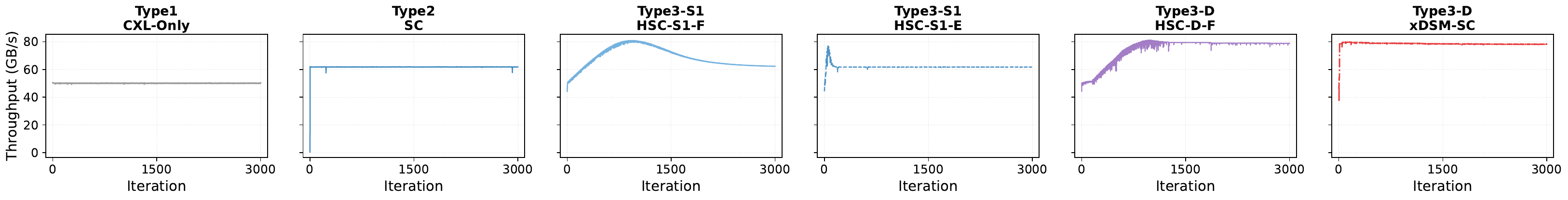}
  \caption{Per-iteration throughput for the iterative program (private read-write, 4.1 GB, 16 threads / 2 nodes).}
  \label{fig:breakdown-loopnest}
\end{figure*}

\begin{figure*}[t]
  \centering
  \Description{Per-iteration TEPS breakdown for BFS (shared read-only intensive, 16 threads).}
  \includegraphics[width=\textwidth]{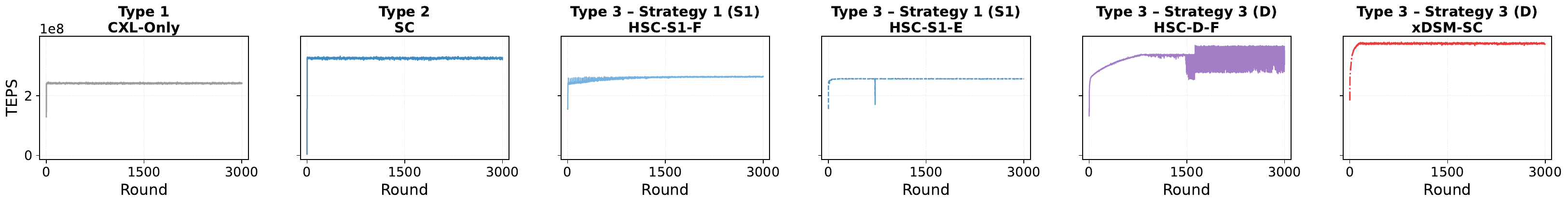}
  \caption{Per-round TEPS (traversed edges per second) for BFS (shared read-only, 7.46 GB, 16 threads / 2 nodes).}
  \label{fig:breakdown-bfs}
\end{figure*}


To understand the effects of \emph{different} configurations, 
we examine per-iteration throughput on two microbenchmarks with complementary access patterns: an iterative program (private read-write, results in Figure~\ref{fig:breakdown-loopnest}) and BFS (shared read-only, results in Figure~\ref{fig:breakdown-bfs}).
{Together, these two patterns cover the dominant access behaviors across all five applications.}
Both run 3\,000 iterations (rounds) on 2 nodes with 16 threads.

\noindent\textbf{Iterative program: private read-write.}
Each thread reads and writes a disjoint partition of a large array, separated by barriers.
We select six configurations for analysis (Figure~\ref{fig:breakdown-loopnest}).
Type\,2 systems behave alike (DRust-C differs only in a faster first iteration), so we show only SC.
For Type\,3, the two consistency protocols (sequential and release consistency) show no difference, so we use sequential consistency.
{\color{blue}HSC-S1 and HSC-S2 perform the same because all data is private, so both strategies promote all pages to local DRAM. We show only HSC-S1.}

\emph{Type\,1 (CXL-ONLY)} delivers a flat ${\sim}$50\,GB/s.
All data resides on CXL, so there is no warm-up, no access to faster local DRAM.
\emph{Type\,2 (SC)} starts at $<$1\,GB/s because the array is initialized on the main thread's node, so worker threads on the other node trigger page faults on every first access.
After ${\sim}$50 iterations, pages settle into local DRAM and the throughput stabilizes at ${\sim}$62\,GB/s, faster than CXL-ONLY but capped by local DRAM bandwidth.


{\color{blue}
\emph{Type\,3, strategy comparison (\textbf{S1/S2 vs.\ D}).}
We fix elastic granularity and compare HSC-S1-E (\textbf{S1}) against \sysname{}-SC (\textbf{D}).
HSC-S1-E peaks at ${\sim}$77\,GB/s around iteration~50 when pages are split between CXL and local DRAM, then drops to ${\sim}$62\,GB/s as \textbf{S1} pulls all pages to local memory, losing the additional CXL bandwidth.
\sysname{}-SC reaches ${\sim}$79\,GB/s by iteration~50 and sustains ${\sim}$78\,GB/s throughout all 3,000 iterations, 27\% above HSC-S1-E at steady state.
This advantage stems from strategy \textbf{D} maintaining a balanced split between CXL and local DRAM, exploiting bandwidth from both tiers.

\emph{Type\,3, granularity comparison (fixed vs.\ elastic).}
We fix strategy D and compare HSC-D-F against \sysname{}-SC.
HSC-D-F starts at ${\sim}$51\,GB/s and slowly climbs to ${\sim}$79\,GB/s by iteration~1\,000, because each page is tracked and migrated individually at 4\,KB granularity.
\sysname{}-SC reaches the same ${\sim}$79\,GB/s by iteration~50, converging $20\times$ faster because elastic pages coalesce contiguous regions (including unaccessed pages) into single management units, migrating up to 512 pages at once.
Both converge to 27\% above Type\,2 and 58\% above CXL-ONLY, confirming that elastic granularity accelerates warm-up without affecting placement quality.
}

\noindent\textbf{BFS: shared read-only.}
Multiple threads perform BFS on a shared CSR graph.
Each round launches all threads in parallel, and each thread traverses from a distinct root.
The graph arrays are read-only and accessed with irregular, pointer-chasing patterns.
Figure~\ref{fig:breakdown-bfs} reports TEPS (traversed edges per second) per round.
Type\,2 systems behave alike, so we show only SC.
HSC-S2 performs the same as HSC-D in this workload, so we show only HSC-D.

\emph{Type\,1 (CXL-ONLY)} is flat at ${\sim}2.4\times 10^{8}$ TEPS.
All graph data sits on CXL, so every pointer chase pays CXL latency.

\emph{Type\,2 (SC)} reaches ${\sim}3.3\times 10^{8}$ TEPS at steady state, 37\% above CXL-ONLY.
The graph is read-only, so SC replicates it into each node's local DRAM without write invalidations.

{\color{blue}
\emph{Type\,3, strategy comparison (S1 vs.\ S2/D).}
We fix elastic granularity and compare HSC-S1-E against \sysname{}-SC.
HSC-S1-E stays flat at ${\sim}2.6\times 10^{8}$ TEPS, nearly identical to CXL-ONLY, because S1 classifies the graph as ``shared'' and keeps it on CXL, unable to recognize that shared-read pages would benefit from local replication.
\sysname{}-SC reaches ${\sim}3.7\times 10^{8}$ TEPS by round~100 and stays stable, 47\% above HSC-S1-E.
Both S2 and D detect that the graph pages are read-only and replicate them to local DRAM, reducing access latency without incurring write invalidations.
This also explains why HSC-S2 performs the same as HSC-D in this workload.

\emph{Type\,3, granularity comparison (fixed vs.\ elastic).}
We fix strategy D and compare HSC-D-F against \sysname{}-SC.
HSC-D-F climbs to ${\sim}3.3\times 10^{8}$ TEPS by round~1\,000, but after round~1\,500 its throughput oscillates between $2.7\times 10^{8}$ and $3.7\times 10^{8}$ TEPS, because at 4\,KB granularity the profiler is more likely to misclassify shared pages as private, causing pages to migrate back and forth between nodes.
\sysname{}-SC reaches ${\sim}3.7\times 10^{8}$ TEPS by round~100 and stays stable, 16\% above Type\,2.
Elastic pages coalesce the shared graph into large management units, providing more accurate page classification and eliminating the oscillation seen in HSC-D-F.
}

\noindent\textbf{Summary.}
{\color{blue}Strategy D adapts placement to access patterns and maintains stable performance via latency-driven migration, outperforming static rules in both workloads.
Elastic granularity enables faster migration and more accurate page classification than fixed pages.
\sysname{} combines both for the best throughput and fastest convergence.}


\subsection{Dynamic Workload}
\label{sec:eval-dynamic}

\begin{figure}[t]
  \centering
  \Description{TPOT over time for LLM inference on Qwen3-30B-A3B driven by Azure trace.}
  \includegraphics[width=\linewidth]{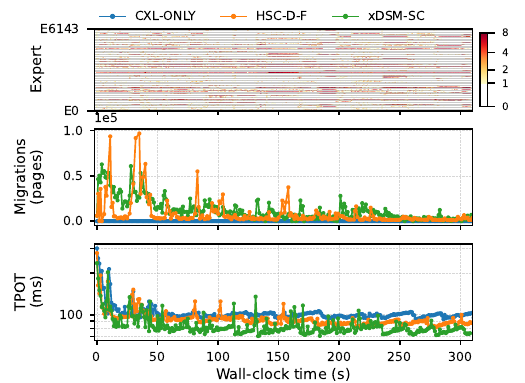}
  \caption{LLM inference (Qwen3-30B-A3B) driven by Azure traces.
  Top: expert activation heatmap, where darker color indicates {\color{blue}higher activation rate}.
  Middle: page migration per second.
  Bottom: TPOT over wall-clock time.}
  \label{fig:dynamic-tpot}
\end{figure}


We run LLM inference using Qwen3-30B-A3B~\cite{qwen3}, a Mixture-of-Experts (MoE) model with 128 experts per layer.
{Each token activates only a small subset of experts, and the activated subset changes across tokens and requests, making the memory access pattern inherently dynamic.}
The int8-quantized model is approximately 37\,GB, exceeding the 16\,GB aggregate local DRAM of four nodes and requiring CXL memory.
We replay a sequence of requests from the Azure LLM Inference Trace~\cite{azurellmtrace} serially on 4 nodes.
Figure~\ref{fig:dynamic-tpot} shows the causal chain from top to bottom.
The top panel shows which experts are activated over time (each row is one expert, darker means more activations). Tracing any single row horizontally, an expert is not activated continuously but intermittently, meaning the working set of weight matrices shifts over time.
The middle panel shows how each system responds with page migration, and the bottom panel shows the resulting Time Per Output Token (TPOT): a system that migrates fast enough to track the shifting activations achieves lower TPOT.


\noindent\textbf{TPOT.}
CXL-ONLY keeps all weights on CXL, yielding a flat TPOT of 101.7\,ms.
Both HSC-D-F and \sysname{}-SC batch-migrate hot expert pages to local memory, but differ in granularity.
HSC-D-F migrates at fixed granularity, reaching 92.6\,ms ($1.10\times$ faster).
\sysname{}-SC migrates at elastic granularity, reaching 82.8\,ms ($1.23\times$ faster than CXL-ONLY, $1.12\times$ faster than HSC-D-F).
Expert weight matrices are large contiguous regions. Elastic pages continuously migrate pages with lower per-page overhead than fixed granularity. Thus, \sysname{}-SC migrates 3.6\,M pages in total versus 1.9\,M for HSC-D-F (Figure~\ref{fig:dynamic-tpot}, middle), 
{adapting to activation shifts faster}.

\noindent\textbf{Goodput.}
Following~\cite{agrawal2024sarathi,zhong2024distserve}, we set a TPOT SLA of 100\,ms and define goodput as the number of output tokens that meet this SLA per second.
CXL-ONLY meets the SLA for only 58.7\% of tokens, yielding a goodput of 5.77\,tok/s.
HSC-D-F reaches 90.5\% SLA attainment and a goodput of 9.77\,tok/s.
\textsc{xDSM}-SC reaches 91.1\% and a goodput of 11.00\,tok/s, $1.13\times$ higher than HSC-D-F and $1.91\times$ higher than CXL-ONLY.

\subsection{Placement Policy Comparison}
\label{sec:eval-policy}

\begin{figure}[t]
  \centering
  \Description{Comparison of alternative placement policies against \sysname{}-SC on 4 nodes (32 threads).}
  \includegraphics[width=\linewidth]{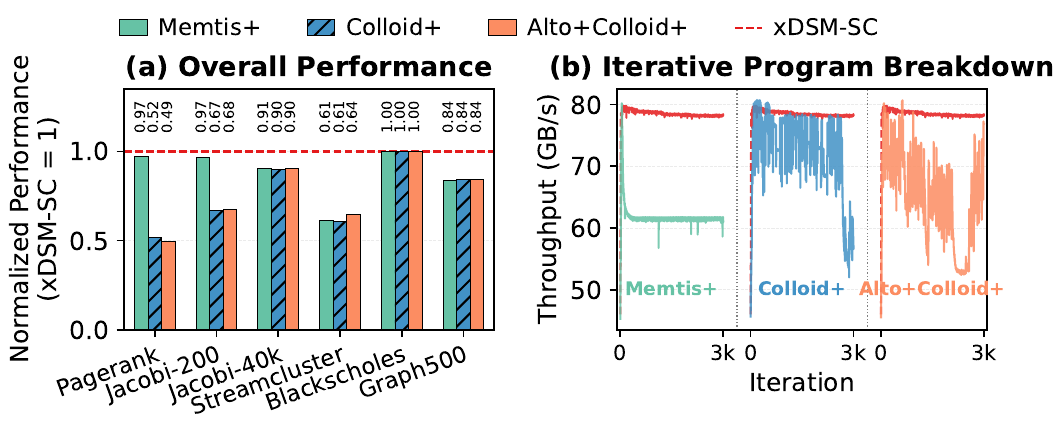}
  \caption{Placement policy comparison on 4 nodes (32 threads).
  (a)~Normalized performance relative to \sysname{}-SC (higher is better).
  (b)~Per-iteration throughput breakdown for the iterative program (3,000 iterations).}
  \label{fig:policy-32t}
\end{figure}

\begin{figure*}[t]
  \centering
  \Description{Scalability results from 1 to 4 nodes across five benchmarks.}
  \includegraphics[width=\textwidth]{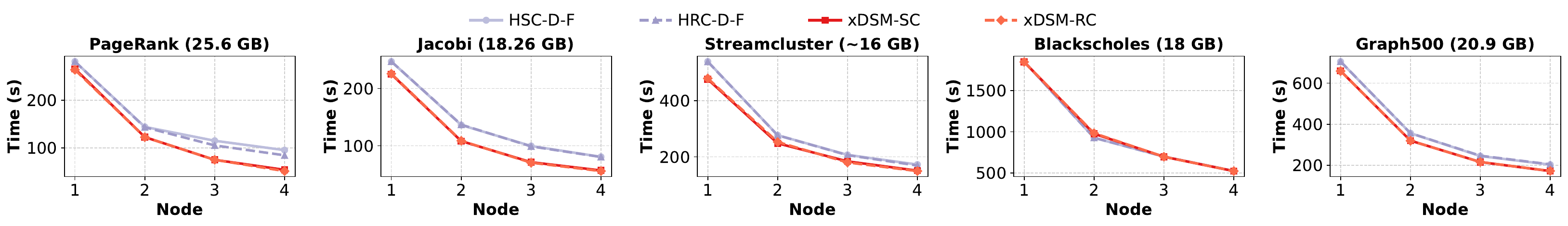}
  \caption{Scalability from 1 to 4 nodes (8 to 32 threads) across five benchmarks with large datasets.
}
  \label{fig:scalability}
\end{figure*}


We compare \sysname{}-SC against three state-of-the-art placement policies adapted to our batch migration runtime. \\ \textbf{Memtis+}~\cite{lee2023memtis} leverages a page access histogram to promote the hottest 2\,MB huge pages locally. \textbf{Colloid+}~\cite{vuppalapati2024colloid} utilizes a binary search algorithm to balance average loaded access latency between memory tiers. Finally, \textbf{Alto+Colloid+}~\cite{liu2025alto} builds upon Colloid+ by integrating Amortized Offcore Latency to adjust migration intensity according to memory-level parallelism.

Figure~\ref{fig:policy-32t} shows the results.
Memtis+ is competitive on short, private-RW-dominated benchmarks (97\% on PageRank and Jacobi-200), because its 2\,MB huge pages adapt quickly.
On longer workloads, Memtis+ greedily fills local DRAM, losing CXL bandwidth (Figure~\ref{fig:policy-32t}b), 10\% slower than \sysname{}-SC on Jacobi-40K.
Memtis+ also lacks shared page management, $1.63\times$ slower on Streamcluster and $1.19\times$ on Graph500.
Colloid+ and Alto+Colloid+ rely on the \emph{average} loaded access latency to guide placement, reaching only 49--52\% on PageRank and 67--68\% on Jacobi-200.
Their binary-search algorithm migrates few pages per step when the latency gap is large, converging slowly.
The average latency also fluctuates between ticks, causing pages to migrate back and forth.
\sysname{} addresses these with an exponential budget that scales migration intensity with the latency gap, and per-node P90 latency histograms that are less sensitive to latency fluctuations.
Like Memtis+, Colloid+ and Alto+Colloid+ also lack shared page management.
Blackscholes is compute-bound and all configurations perform within 1\% variance.

\subsection{Scalability}
\label{sec:eval-scalability}

We run all five benchmarks with \emph{large} datasets (16--25.6\,GB) that exceed the aggregate local DRAM of all four nodes (16\,GB).
\textbf{At this scale, Type\,2 local-only DSM systems \emph{cannot run}}: the working set does not fit in local DRAM even when all nodes participate.
The hybrid configurations (Type\,3), which leverage the 64\,GB CXL shared memory pool, can execute these workloads.
Figure~\ref{fig:scalability} shows performance from 1 node (8 threads) to 4 nodes (32 threads).

\noindent\textbf{Near-linear scaling.}
The memory-intensive benchmarks scale well: PageRank reaches $5.1\times$, while Jacobi reaches $4.0\times$, Graph500 $3.8\times$, and Streamcluster $3.2\times$.
Blackscholes, being compute-bound, also achieves $3.5\times$ speedup at 4 nodes.
{\color{green}
PageRank achieves superlinear scaling ($5.1\times$ on 4 nodes) by leveraging increased aggregate local DRAM. With a 25.6\,GB dataset, a single node's 4\,GB memory forces most data onto the CXL tier. Expanding to 4 nodes provides 16\,GB of local capacity, shifting data locally and fully utilizing local bandwidth. Consequently, steady-state throughput jumps massively from ${\sim}$21\,GB/s on 1 node to ${\sim}$117\,GB/s on 4 nodes.
}
{Additionally, the elastic variants (\sysname{}-SC/RC) outperform their fixed-page counterparts (HSC/HRC-D-F), and the advantage grows with node count: at 4 nodes, elastic groups reduce execution time by 39\% for PageRank, 30\% for Jacobi, 16\% for Graph500, and 12\% for Streamcluster. Blackscholes performs within 0.3\% across all configurations.}


%

\section{Discussion}
\label{sec:dis}


{The performance gains from dynamic page placement are fundamentally tied to the latency disparity between local and remote memory. Our NUMA-based emulation exhibits a 2$\times$--4$\times$ latency gap, under which our policy already demonstrates substantial efficacy. In actual CXL 3.0 deployments, remote accesses traverse CXL switches~\cite{cxl30whitepaper}, introducing network latencies absent in our emulation. Because a wider latency gap penalizes remote accesses more severely, the benefits of promoting hot pages to local memory will increase on physical CXL hardware.}

Currently, the \texttt{eptp\_database} resides in globally shared CXL memory, operating under a trusted-environment assumption. Consequently, a compromised node could maliciously read or tamper with the mapping metadata of other nodes. Hardening this shared state against adversarial access is left as future work, which could be addressed by integrating hardware-enforced memory protection mechanisms, such as Memory Protection Keys (MPK)~\cite{intelsdm-mpk}.

\section{Related Work}
\label{sec:related}

\noindent\textbf{Transparent Execution Environment.}
GiantVM~\cite{zhang2020giantvm} and vSMP Foundation~\cite{auburn_vsmp} aggregate multiple physical machines into a single virtual machine to enable the execution of unmodified applications across multiple nodes. Fundamentally, these systems still rely on traditional software consistency protocols~\cite{li1988ivy}. Consequently, they share the same performance limitations as the Type 2 systems evaluated in our study (\S\ref{sec:eval-e2e}). Furthermore, our analysis reveals that in sub-$\mu\mathrm{s}$ CXL environments, the specific choice of software-based consistency protocol~\cite{li1988ivy,keleher1994lazy,zhou1996hlrc,mueller1997dsmthreads,nelson2015grappa,bennett1990munin,bershad1993midway,scales1997shasta,kim2020dex,ma2024drust} has a \emph{limited impact on overall performance}.
MIND~\cite{lee2021mind} also provides a global shared address space, but builds it on a different hardware substrate, an in-network programmable switch, rather than the CXL-attached shared memory that \sysname{} targets.



\noindent\textbf{Page Placement Policies.}
Tiered memory systems (\textbf{Type 3}) allow direct access to both local and remote (CXL) memory. They migrate pages either based on frequency~\cite{maruf2023tpp,raybuck2021hemem,lee2023memtis,xiang2024nomad,Duraisamy2023TMTS,zhong2024memstrata} or latency~\cite{vuppalapati2024colloid,liu2025alto}. Frequency-based systems identify hot pages and try to promote them to local memory but may lead to local contention. Latency-based systems avoid such contention. \sysname{} improves these policies by using the  exp-binned histograms and maintaining a latency equilibrium between local and remote memory(\S\ref{sec:eval-policy}). Far memory systems (\textbf{Type 2})~\cite{gu2017infiniswap,guo2023mira,chen2024atlas,tauro2024trackfm,zhong2024unimem,li2025beehive} use remote memory solely as a backing tier, forcing all accessed pages into local memory before execution. When applied to CXL, this approach squanders the ability to access local and remote memory simultaneously. \sysname{} operates as a Type 3 system to avoid this limitation.

\noindent\textbf{Adaptive Sharing Granularity.}
While \textsc{AdaptableView}~\cite{itzkovitz2000adaptableview} also recognizes the performance impact of sharing granularity, it relies heavily on manual programmer annotations to define sharing views for specific variables and execution phases.
In contrast, \sysname{} demonstrates a highly feasible automated solution. By transparently tracking runtime access patterns, \sysname{} dynamically adapts the management granularity on the fly, completely eliminating the need for source code modifications.
MIND~\cite{lee2021mind} also adopts variable-sized \emph{regions}, but they define the granularity of its coherence directory, while data is transferred at a fixed 4\,KB page. In \sysname{}, the elastic page is itself the unit of transfer, coalesced and split by access pattern and consistency state.

\noindent\textbf{Other CXL-based Systems.}
Pond~\cite{pond2023} pools CXL memory to avoid stranded capacity, focusing on allocation strategy and performance prediction.
Telepathic~\cite{mahar2024telepathic} shares native pointers across nodes to bypass serialization, focusing on memory access isolation.
TrEnv~\cite{huang2024trenv} and CXLfork~\cite{alverti2025cxlfork} share process snapshots across nodes, focusing on page table manipulation and copy-on-write.
Tigon~\cite{huang2025tigon} synchronizes transactions on CXL shared memory, focusing on locking protocols co-designed with coherence boundaries.
\sysname targets general-purpose distributed shared memory over CXL, focusing on page placement and elastic granularity.
\section{Conclusion}

\sysname{} seamlessly scales multi-threaded applications across CXL-connected nodes. By co-designing a kernel module and user-space runtime, it delivers a full-space shared environment, latency-driven dynamic page placement, and elastic page management. Evaluations show that \sysname{} outperforms existing systems, adapts to dynamic workloads, and achieves near-linear scalability.


\bibliographystyle{ACM-Reference-Format}
\bibliography{dthreads-refs}

\end{document}